\documentclass[journal=jacsat,manuscript=article]{achemso}

\author{David C. Malaspina}
\affiliation{Fundacio Universitat Rovira i Virgili, Av. dels Paisos Catalans 18, 43007, Tarragona, Spain}

\email{davidcesar.malaspina@fundacio.urv.cat}
\author{Jordi Faraudo}
\affiliation{Institut de Ciencia de Materials de Barcelona (ICMAB-CSIC), Campus UAB, E-08193 Bellaterra, Barcelona, Spain.}
\email{jfaraudo@icmab.es}
\title[An \textsf{achemso} demo]
  {Comment on "Nanoscale Wetting of Crystalline Cellulose"}

\abbreviations{IR,NMR,UV}

\begin{document}

%\begin{tocentry}

%Some journals require a graphical entry for the Table of Contents.
%This should be laid out ``print ready'' so that the sizing of the

%Inside the \texttt{tocentry} environment, the font used is Helvetica
%8\,pt, as required by \emph{Journal of the American Chemical

%The surrounding frame is 9\,cm by 3.5\,cm, which is the maximum
%permitted for  \emph{Journal of the American Chemical Society}
%graphical table of content entries. The box will not resize if the
%content is too big: instead it will overflow the edge of the box.

%This box and the associated title will always be printed on a
%separate page at the end of the document.

%\end{tocentry}

\begin{abstract}
In a recent publication, Trentin et al.\cite{Munir} employed Molecular Dynamics (MD) simulations for the theoretical study of the wetting of the different polymorphs of cellulose by water, using the widely employed TIP3P model of water. Here we show that the selection of the particular water model employed in the simulations has a critical impact in the results, a point overlooked by the authors. In particular, the TIP3P model of water has an unrealistically low value of the surface tension which compromises wetting studies made with this model. Slightly more complex models such as TIP4P2005 correctly reproduce the surface tension of water. As a consequence, the results of MD simulations of cellulose wetting  using the low-tension TIP3P model show full wetting in situations that the more realistic TIP4P2005 water model predicted the formation of a water droplet onto cellulose.
\end{abstract}

In a recent work, Trentin et al. \cite{Munir} make an extensive theoretical study of the wetting behaviour of different cellulose surfaces from different cellulose crystal polymorphs and different crystal planes, using Molecular Dynamics (MD) simulations.
Some of their results are not fully consistent with our previous study  \cite{Malaspina} in which we considered the interaction of two different cellulose surfaces with different solvents and different molecules.
The reason of the discrepancies with earlier work were not identified by Trentin et al. \cite{Munir} but, as shown here, they can easily traced back to the particular water model employed in the simulations.
In their work, Trentin et al. considered the highly simplified TIP3P water model, which predicts an unrealistically low value of the surface tension of water of $\gamma_{lv}=$52.3 mN/m (see Table III in Ref \cite{Tension}), which should be compared with the experimental value of $\gamma_{lv}=$71.7 mN/m. 
The surface tension of water $\gamma_{lv}$ plays a fundamental role in determining the contact angle $\theta$ of a water droplet onto a solid surface, as dictated by Young's equation:
\begin{equation}
\label{Young}
\gamma_{lv}cos(\theta)=\gamma_{sv}-\gamma_{sl},
\end{equation}
where $\gamma_{sv}$ and $\gamma_{sl}$ are the solid-vapour and solid-liquid surface tensions, respectively.
Eq.(\ref{Young}) shows that wetting simulations using the low-tension TIP3P water model will predict much lower contact angles than those expected from the actual, higher surface tension of water. 
For this reason, TIP3P is not a suitable model for simulations involving the surface tension of water, and in particular, wetting simulations. 

Since we were aware of these defects of the popular TIP3P model of water, in our previous work \cite{Malaspina} we considered the more realistic (and computationally more expensive) TIP4P2005 model. This model predicts a surface tension for water of $\gamma=$69.3 mN/m (see Table IV in Ref \cite{Tension}), close to the experimental result (71.7 mN/m) so it is therefore a model suitable for wetting simulations.

\begin{figure}[ht]
  \includegraphics[width=1.0\columnwidth]{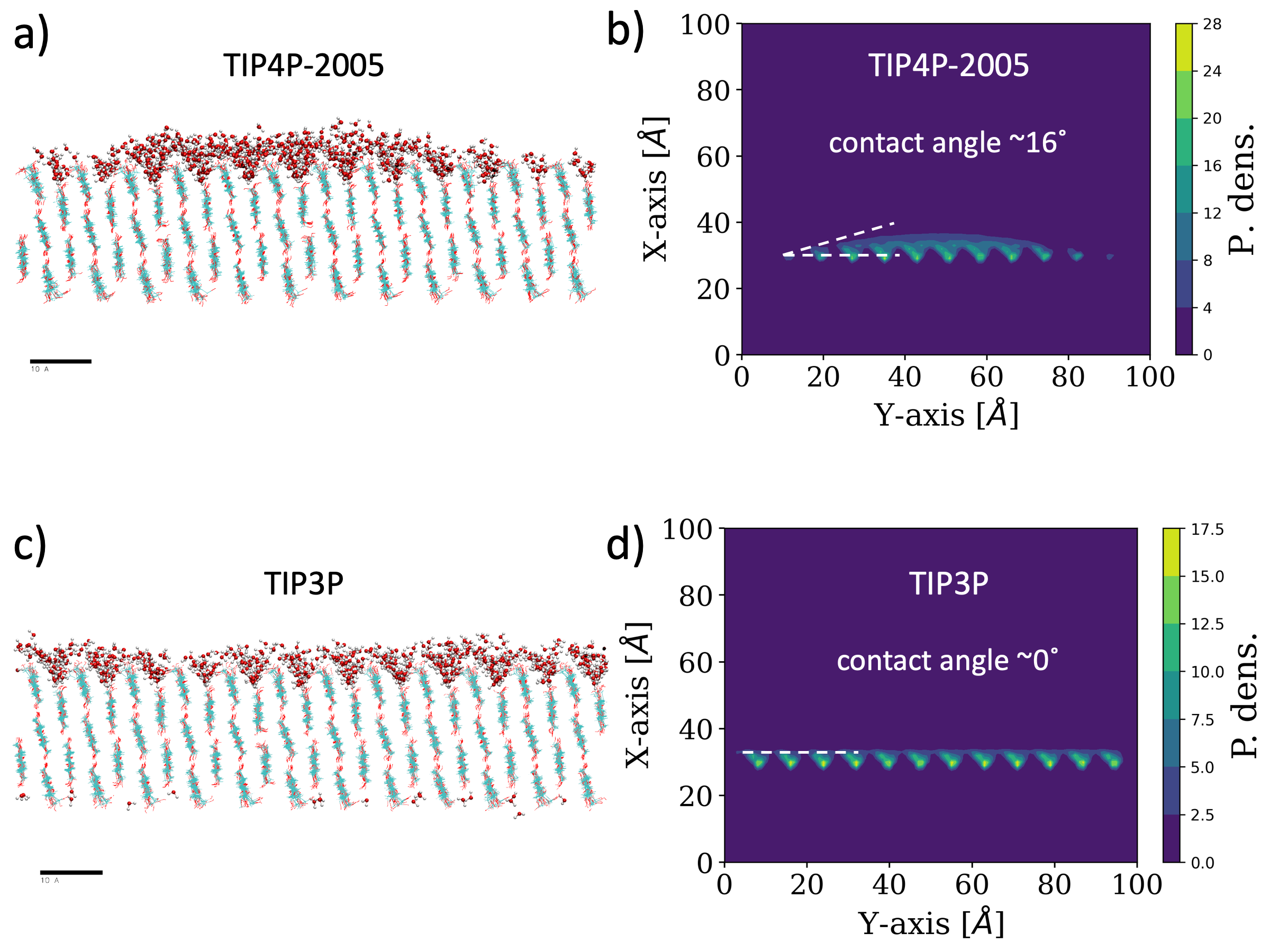}
  \caption{Results of MD simulations of wetting of plane (010) of I$\beta$ cellulose with different models of water. The results shown in a) and b) correspond to TIP4P-2005 water and the results shown in c) and d) correspond to TIP3P water model. Panels a) and c) show snapshots of the simulation. Cellulose is shown as lines and water molecules are shown in CPK representation. Panels b) and d) show the 2-D density profile of water molecules. The resulting contact angles are also indicated.  }
  \label{fgr:fig1}
\end{figure}

In order to show more clearly the effect of the particular water model employed in the calculations, we have repeated one of our previous MD simulations \cite{Malaspina} changing the water model.
We considered the particular case of the wetting of the cellulose plane (010) of I$\beta$ cellulose.
All the details of the new calculation (simulation parameters, simulation time, number of water molecules, ...) are the same as the original one reported in Figure 3 in Ref. \cite{Malaspina} with the only difference being the water model (the original TIP4P2005 water model\cite{Malaspina} is now replaced by the TIP3P model).
The results are shown in Figure 1.
As shown in this figure, the water droplet observed over cellulose (with contact angle of approximately 16º) in the case of the TIP4P2005 water model transforms into full wetting in the case of simulations with the TIP3P water.
In other words, the low surface tension of TIP3P is not enough to sustain a droplet onto this cellulose surface.

This result emphasizes the strong dependence of wetting results on the particular model of water and highlights the necessity of using water models with realistic values of the water liquid-vapour surface tension in wetting simulations.

\begin{acknowledgement}

This work was supported by the Spanish Ministry of Science
and Innovation through Grant No. RTI2018-096273-B-I00, and the
‘‘Severo Ochoa’’ Grant No. CEX2019-000917-S for Centres of
Excellence in R\&D awarded to ICMAB

\end{acknowledgement}

%\bibliography{references}

\end{document}